\documentclass[11pt]{article}
\usepackage{graphicx,amssymb,amsmath}
\usepackage{bibentry} 
\usepackage{multicol} 

\usepackage{setspace} 
\usepackage[version=4]{mhchem}
\usepackage{cite}

\usepackage{authblk}
\usepackage{xcolor}

\newcommand{\br}{{\mathbf{r}}}
\newcommand{\bu}{{\mathbf{u}}}
\newcommand{\bxi}{{\boldsymbol{\xi}}}
\newcommand{\be}{{\mathbf{e}}}

\begin{document}

\title{Active-Passive Brownian Particle in Two Dimensions}
\author {Tayeb Jamali\footnote{Email: tjamali.official@gmail.com}}
\affil{\small{Broad Institute of MIT and Harvard, Cambridge, MA, USA}}

\maketitle \baselineskip 24pt

\singlespacing

\begin{abstract}

We describe a two–dimensional model for active particles whose self–propulsion speed is not fixed, but varies in time, and whose motion is subject to both translational and rotational diffusion. In the conventional treatment of active Brownian motion, the self–propulsion speed is taken to be constant — an assumption convenient for analysis but poorly matched to many real systems. Here we relax that assumption, allowing the speed $v(t)$ to fluctuate stochastically between two values: $v=0$ (a passive state) and $v=s$ (an active state). Transitions between these states are taken to follow a random telegraph process. This “active–passive Brownian particle” inherits limiting behaviors from both the purely active and purely passive Brownian cases. Analytical expressions for the first two displacement moments, and for the resulting effective diffusion coefficient, make this dual character explicit. Moreover, by an appropriate identification of parameters, a run–and–tumble particle — such as a motile bacterium — can be mapped onto this model in such a way that their large–scale diffusivities coincide.

\end{abstract}

\section{Introduction}

What does it mean for a particle to move actively only some of the time?
In the simplest theories of active matter, self-propulsion is taken to be a permanent condition — the particle is always “on,” pushing itself forward at a fixed speed. Such an assumption makes the mathematics tidy and the physical picture simple. But many real systems are less obliging: their motion is a stop-and-go affair, with periods of vigorous activity punctuated by moments of passivity. The challenge, then, is to understand how this intermittent activity reshapes the statistical properties of motion, and how to capture both limits — the always-on and the never-on — in a single framework.

Cellular motility plays an essential role in biological processes \cite{Bray_2000}. Wound healing, for example, cannot proceed without the coordinated movement of various types of cells, such as fibroblasts \cite{Velnar_2009}. Nor is this phenomenon confined to the cells of multicellular organisms: unicellular microorganisms, including bacteria, are able to crawl along surfaces or swim through fluid environments \cite{Berg_1975,Harshey_2003,Jarrell_2008}.

A characteristic feature of such motion is its variability in both speed and direction over time. These variations may arise from stochastic fluctuations at the cellular scale, or from deliberate changes in response to environmental cues. The latter case is familiar from studies of bacterial chemotaxis, in which microorganisms adjust their trajectories according to chemical or physical stimuli in their surroundings \cite{Dusenbery_2011,Adler_1966}. Whatever their origin, temporal changes in self-propulsion speed are not a mere complication to be averaged away: they are central to the mechanics of motile cells, and to understanding the many biological mechanisms whose operation depends on that motility.

The term \emph{active particle} (or \emph{self-propelled particle}) is applied broadly to microscopic entities — from living cells and microorganisms to synthetic particles — that are capable of converting energy from their surroundings into directed motion \cite{Schimansky_1995,Ramaswamy_2010,Marchetti_2013}. Such particles are intrinsically out of equilibrium. Their conceptual simplicity, combined with the rich range of behaviors they exhibit, has made them an attractive subject for both experimental and theoretical study \cite{Howse_2007,Lauga_2009,Walther_2013,Elgeti_2015,Volpe_2016,Zottl_2016}.

Among the theoretical frameworks developed to describe active motion, \emph{active Brownian motion} (ABM) is perhaps the most widely used \cite{Schweitzer_2003}. It extends the classical Brownian motion model by adding a constant-magnitude self-propulsion force and orientational diffusivity for the direction of motion (see \cite{Romanczuk_2012} and references therein). While the assumption of constant self-propulsion speed is reasonable for many applications, it is by no means universal. In reality, self-propulsion can vary in time — sometimes dramatically — and several studies have addressed the consequences of such variability. Peruani \textit{et al.} \cite{Peruani_2007} analyzed fluctuations in both speed and direction, identifying characteristic timescales and distinct dynamical regimes. Babel \textit{et al.} \cite{Babel_2014} examined active Brownian motion with three deterministic forms of time-dependent self-propulsion speed, and characterized the resulting statistical properties.

Here we take a different approach. We retain the active Brownian motion framework but allow the self-propulsion speed to vary stochastically in time. Specifically, the speed $v(t)$ alternates between an “active” state ($v=s$) and a “passive” state ($v=0$), with the switching governed by a random telegraph process. This \emph{active–passive Brownian particle} model permits exact analytical results for the first two displacement moments, from which an effective diffusion coefficient can be obtained. The analysis reveals that the model combines, in a single description, the asymptotic behaviors of both standard Brownian motion and active Brownian motion.

As an application, we show that the large–scale diffusivity of a non-interacting run–and–tumble particle (RTP) — a well-known model for motile bacteria such as \emph{Escherichia coli} — can be matched exactly by an appropriately parameterized active–passive Brownian particle. Thus, the present model offers a compact, analytically tractable framework for systems in which self-propulsion is effectively “switched on and off” in time.

\section{Model and Results}

\begin{figure}[t!]
\centering
\includegraphics[width=0.3\textwidth]{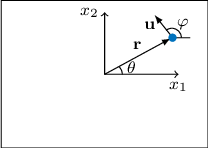}
\caption{A schematic active–passive Brownian particle: the small blue disk is at position $\br$ and moves in the unit direction $\bu$ set by the orientation angle $\varphi$ relative to the $x_1$-axis.}
\label{fig: ABP in plane}
\end{figure}

We begin with a particle at position $\br = (x,y)$, whose self-propulsion speed $v(t)$ varies in time, moving in the direction $\bu(\varphi)=(\cos\varphi,\sin\varphi)$, and subject to both translational and rotational diffusion. A schematic of this setup appears in Fig.~\ref{fig: ABP in plane}. The motion is governed by the two-dimensional Langevin equations
\begin{align}
\label{eq: translational Langevin eq}
\dot{\br} &= v(t) \,\bu(\varphi) + \sqrt{2 D_t}\,\bxi(t),\\ 
\label{eq: rotational Langevin eq}
\dot{\varphi} &= \sqrt{2D_r}\,\eta(t),
\end{align}
where $\bxi(t) = (\xi_x(t),\xi_y(t))$ and $\eta(t)$ are independent Gaussian white noises of zero mean and unit variance. These random terms express the local thermal forces and torques from the surrounding medium. The constants $D_t$ and $D_r$ are the translational and rotational diffusion coefficients, respectively.  

If $v(t)$ were constant, Eqs.~(\ref{eq: translational Langevin eq})--(\ref{eq: rotational Langevin eq}) would describe the standard active Brownian particle (ABP). Here we allow $v(t)$ to take only two possible values, $0$ and $s$, switching randomly between them. In other words, the particle alternates between an \emph{active} state and a \emph{passive} state. We will refer to this as the \emph{active–passive Brownian particle} model.  

The two-state choice for $v(t)$ is meant to capture a familiar biological fact: many microorganisms are not continuously motile but pass through intervals of inactivity. We model the switching as a Markov process with constant rates
\begin{equation}
\label{eq: v(t)}
v(t): 0 \ce{<=>[\alpha][\beta]} s,
\end{equation} 
where $\alpha$ is the activation rate and $\beta$ the deactivation rate. This is the well-known \emph{random telegraph process}—simple enough for analytical work but still rich enough to capture the essentials of on–off self-propulsion.

To see what this model implies, we look at the first two displacement moments, $\langle \br(t)-\br_0 \rangle \quad \text{and} \quad \langle (\br(t)-\br_0)^2 \rangle$, with $\br_0\equiv\br(t=0)$. The second moment is the mean-square displacement (MSD). It is convenient to work component-wise. Writing Eq.~(\ref{eq: translational Langevin eq}) along each Cartesian axis,
\begin{align}
\label{eq: translational Langevin equation 2}
 \begin{split}
  \dot{x}_i &= v(t)\,\be_i \!\cdot\! \bu(\varphi) + \sqrt{2D_t}\,\be_i \!\cdot\! \bxi(t), 
  \quad i=1,2,
 \end{split}  
\end{align} 
with $\be_1=(1,0)$, $\be_2=(0,1)$.  

From here the first and second moments in direction $i$ are
\begin{equation}
 \label{eq: first moment of x}
 \langle x_i(t)-x_i(0) \rangle = \int_0^t  dt_1 \,\langle v(t_1) \rangle \,\langle \be_i \!\cdot\! \bu(\varphi(t_1)) \rangle,
\end{equation} 
\begin{equation}
\label{eq: second moment of x}
  \langle [x_i(t)-x_i(0)]^2 \rangle = 2D_t\, t 
  + \int_0^t\!dt_1 \int_0^t\!dt_2 \, \langle v(t_1) v(t_2) \rangle \,\langle \be_i \!\cdot\! \bu(\varphi(t_1))\, \be_i \!\cdot\! \bu(\varphi(t_2)) \rangle.
\end{equation} 

The factorization here assumes $v(t)$ and $\varphi(t)$ are independent. This reduces the problem to evaluating six basic averages:
\[
\langle v(t) \rangle,\quad \langle v(t_1) v(t_2) \rangle,\quad \langle \cos\varphi(t) \rangle,\quad \langle \sin\varphi(t) \rangle,\quad \langle \cos\varphi(t_1)\cos\varphi(t_2) \rangle,\quad \langle \sin\varphi(t_1)\sin\varphi(t_2) \rangle,
\]
derived in Appendices~\ref{sec: Random Telegraph Process} and~\ref{sec: Angular Ensemble Averages}.  

\begin{figure}[t!]
\centering
\includegraphics[width=0.7\textwidth]{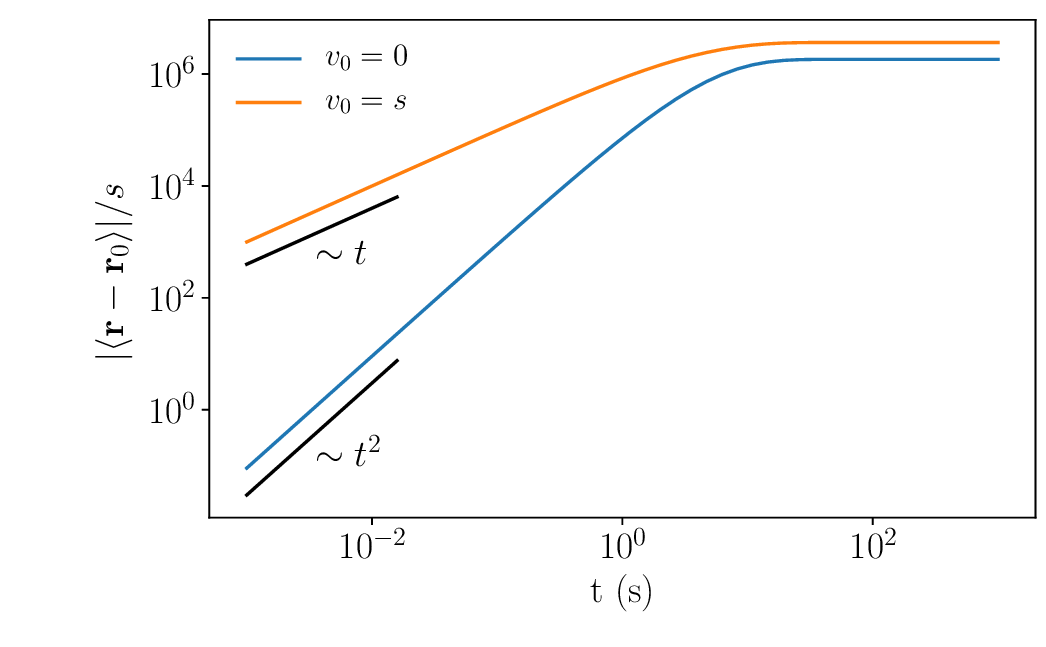}
\caption{Mean displacement $\langle \br(t)-\br_0 \rangle$ as a function of time for a particle of radius $R=1 \,\mu$m in water at room temperature, with $D_t=0.24\, \mu\mathrm{m}^2/\mathrm{s}$, $\alpha=\beta=D_r=0.18\,\mathrm{s}^{-1}$, and $s=3 \,\mu\mathrm{m}/\mathrm{s}$. The two curves correspond to starting from rest ($v_0=0$) and starting active ($v_0=s$). The figure illustrates the short-time quadratic versus linear growth and the eventual saturation predicted by Eq.~(\ref{eq: average of r(t)}).}
\label{fig: MeanPosition}
\end{figure}

Substituting $\langle v(t) \rangle$, $\langle \cos\varphi(t) \rangle$, and $\langle \sin\varphi(t) \rangle$ into Eq.~(\ref{eq: first moment of x}), we obtain
\begin{equation}
 \label{eq: average of r(t)}
 \langle \br(t)-\br_0 \rangle = s\, \bu_0 \left[ \frac{\alpha}{\alpha+\beta} \left( \frac{1-e^{-D_r t}}{D_r} \right) + \left( \delta_{v_0,s} - \frac{\alpha}{\alpha+\beta} \right) \left( \frac{1-e^{-(\alpha+\beta+D_r) t}}{\alpha+\beta+D_r} \right) \right],
\end{equation} 
with $\bu_0 = (\cos\varphi_0,\sin\varphi_0)$ the initial propulsion direction. The two exponentials in Eq.~(\ref{eq: average of r(t)}) already signal distinct temporal regimes.

For $t \ll (\alpha+\beta+D_r)^{-1}$, expansion gives
\begin{equation}
\label{eq: M(near zero)}
\langle \br(t)-\br_0 \rangle = s\,\bu_0 \left[ \frac{1}{2}\alpha\, t^2 + \delta_{v_0,s} \left( t - \frac{1}{2} (\alpha+\beta+D_r) t^2 \right) + O(t^3) \right],
\end{equation}
so a particle starting from rest ($v_0=0$) grows quadratically in time, while one starting active ($v_0=s$) grows linearly.  

For $t \gg D_r^{-1}$,
\begin{equation}
\label{eq: M(infinity)}
\langle \br(t)-\br_0 \rangle \approx s\,\bu_0 \left[ \frac{\alpha}{\alpha+\beta} \frac{1}{D_r} + \left(\delta_{v_0,s}-\frac{\alpha}{\alpha+\beta} \right) \frac{1}{\alpha+\beta+D_r} \right],
\end{equation}
approaching a constant that again depends on $v_0$. This constant depends, once again, on the initial speed $v_0$. To give it shape, we take a disk of radius $R=1\,\mu \mathrm{m}$ suspended in water at room temperature. From the Stokes–Einstein relations, $D_t=k_B T/6\pi\eta R$ and $D_r=k_B T/8\pi\eta R^3$, with $k_B$ the Boltzmann constant and $\eta \approx 0.9 \,\mathrm{mPa}$. We set the active-state speed to $s=3 \,\mu\mathrm{m/s}$ and take $\alpha=\beta=D_r$, values that fall within the range reported for both biological and synthetic particles~\cite{Volpe_2016}. Figure~\ref{fig: MeanPosition} shows the mean displacement for a particle that begins at rest and for one that begins already in the active state. The crossover to the asymptotic forms is evident in both cases.

Continuing with Eq.~(\ref{eq: second moment of x}), and substituting the needed averages, we find the MSD
\begin{equation}
\label{eq: MSD}
\langle (\br(t)-\br_0)^2 \rangle = A_0 + A_1\, t + A_2 e^{-D_r t} + A_3 e^{-(\alpha+\beta)t} + A_4 e^{-(\alpha+\beta+D_r)t},
\end{equation}
with coefficients given as follows:

\begin{align}
\label{eq: MSD's coefficients}
\nonumber
A_0 &=  \frac{s^2}{(\alpha+\beta)^2} \bigg\{-2\frac{\alpha^2}{D_r^2} -2\frac{\alpha\beta}{(\alpha+\beta+D_r)^2} \\ \nonumber 
&\quad+ (\delta_{v_0,s}(\alpha+\beta)-\alpha) \bigg[ \frac{\alpha}{\alpha+\beta-D_r}\left(\frac{1}{D_r}-\frac{1}{\alpha+\beta}\right)  - \frac{\beta}{\alpha+\beta+D_r}\left(\frac{1}{D_r}-\frac{1}{\alpha+\beta}\right) + \frac{1}{D_r} \bigg] \bigg\}, \\  \nonumber
A_1 &=  4D_t + 2 \alpha \frac{s^2}{(\alpha+\beta)^2}  \bigg\{ \frac{\alpha}{D_r} + \frac{\beta}{\alpha+\beta+D_r}\bigg\},\\ \nonumber
A_2 &=  2 \frac{s^2}{(\alpha+\beta)^2} \frac{\alpha}{D_r} \bigg\{ \frac{\alpha}{D_r} - \frac{\delta_{v_0,s}(\alpha+\beta)-\alpha}{\alpha+\beta-D_r}\bigg\},\\ \nonumber
A_3 &=  \frac{s^2}{(\alpha+\beta)^2} (\delta_{v_0,s}(\alpha+\beta)-\alpha) \bigg\{ \frac{\alpha}{\alpha+\beta}\frac{1}{\alpha+\beta-D_r} - \frac{\beta}{\alpha+\beta} \frac{1}{D_r} \\ \nonumber
&\quad - \frac{\alpha}{D_r} \bigg( \frac{1}{\alpha+\beta} - \frac{1}{\alpha+\beta-D_r} \bigg) -\frac{\beta}{\alpha+\beta+D_r} \bigg( \frac{1}{\alpha+\beta} + \frac{1}{D_r} \bigg)\bigg\},  \\ 
A_4 &=  2 \frac{s^2}{(\alpha+\beta)^2} \frac{\beta}{\alpha+\beta+D_r} \bigg\{ \frac{\delta_{v_0,s}(\alpha+\beta)-\alpha}{D_r} + \frac{\alpha}{\alpha+\beta+D_r}\bigg\}.
\end{align} 

\begin{figure}[t!]
\centering
\includegraphics[width=0.7\textwidth]{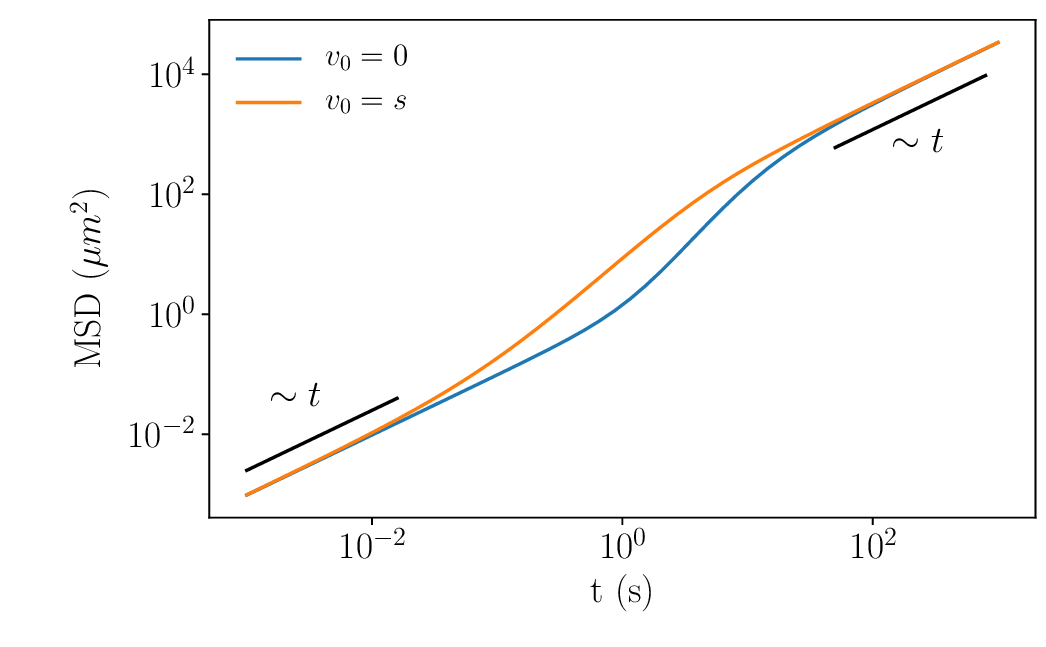}
\caption{Mean-square displacement for the same parameters as in Fig.~\ref{fig: MeanPosition}. At short times, the MSD grows linearly, reflecting the initial diffusive regime. At intermediate times, nonlinear behavior emerges as the interplay between switching dynamics $(\alpha+\beta)^{-1}$ and rotational diffusion $D_r^{-1}$ becomes significant. At long times, the MSD returns to linear growth, now characterized by the effective diffusion coefficient $D_{eff}$ given by Eq.~(\ref{eq: effective diffusion coefficient}).}
\label{fig: MSD}
\end{figure}

The structure of Eq.~(\ref{eq: MSD}) reveals three characteristic time scales: the rotational diffusion time $D_r^{-1}$, the switching time $(\alpha+\beta)^{-1}$, and the combined scale $(\alpha+\beta+D_r)^{-1}$. The longest of these, $D_r^{-1}$, marks the loss of orientational correlations, while the shortest, $(\alpha+\beta+D_r)^{-1}$, reflects the rapid interplay of switching and rotational noise. The intermediate scale $(\alpha+\beta)^{-1}$ arises purely from the switching dynamics. The separation between these scales is determined by the values of $\alpha$ and $\beta$. For times well outside the interval $[(\alpha+\beta+D_r)^{-1},D_r^{-1}]$, the MSD grows linearly in $t$, while within this interval nonlinear behavior can emerge. Depending on parameters, the nonlinear segment may include ballistic growth, MSD $\sim t^2$. To illustrate, we use the same parameters as in Fig.~\ref{fig: MeanPosition}. Figure~\ref{fig: MSD} shows the resulting MSD for particles starting either from rest ($v_0=0$) or with finite initial speed ($v_0=s$), highlighting the crossover between these distinct regimes.

\begin{figure}[t!]
\centering
\includegraphics[width=0.7\textwidth]{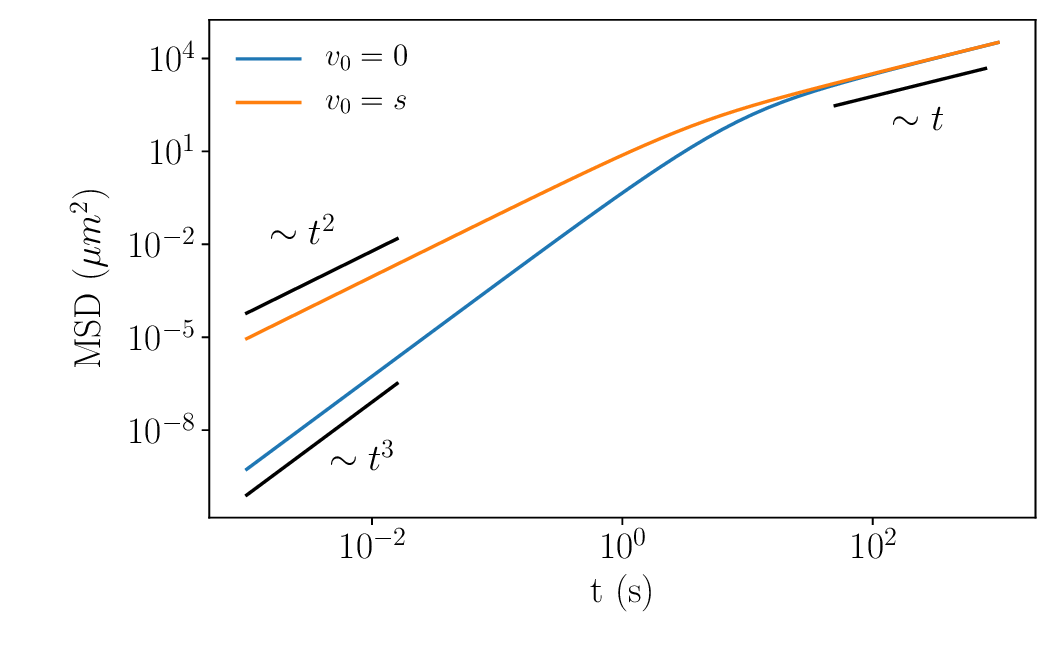}
\caption{Mean-square displacement when $D_t=0$, all other parameters as in Fig.~\ref{fig: MeanPosition}. With translational diffusion suppressed, the short-time linear regime disappears, and intermediate-time ballistic or superdiffusive behavior becomes prominent, as anticipated from the analysis of Eq.~(\ref{eq: MSD}).}
\label{fig: MSD2}
\end{figure}

In many experimental systems, the translational diffusivity $D_t$ is negligible compared to the self-propulsion speed. Wild-type run-and-tumble bacteria are a familiar example, where $D_t$ can safely be taken as zero~\cite{Cates_2013}. In this case the short-time linear regime at $t \ll (\alpha+\beta+D_r)^{-1}$ disappears, leaving ballistic–diffusive and superdiffusive behavior before the eventual linear growth sets in at $t \gg D_r^{-1}$. Figure~\ref{fig: MSD2} illustrates this situation. A particularly simple limit occurs when $v_0=s$ and $\beta=0$: the particle remains in the active state indefinitely, and the dynamics reduce to those of a standard active Brownian particle with constant speed $s$.

Among the terms in Eq.~(\ref{eq: MSD}), only the linear one, $A_1 t$, survives at long times. It defines the effective diffusion coefficient, introduced in general as~\cite{Pottier_2014}
\begin{equation}
D_{eff} = \lim_{t\rightarrow\infty}
\frac{\langle (\br(t) - \br_0 )^2\rangle}{2d\,t},
\end{equation}
with $d$ the spatial dimension. For the active–passive Brownian particle this becomes
\begin{equation}
\label{eq: effective diffusion coefficient}
D_{eff} = D_t + \frac{s^2}{2D_r} \frac{\alpha}{(\alpha+\beta)^2}
\left( \alpha+ \beta \frac{D_r}{\alpha+\beta+D_r}\right).
\end{equation}
When $\beta=0$, Eq.~(\ref{eq: effective diffusion coefficient}) reduces to the well-known ABP result $D_{eff}=D_t+s^2/2D_r$. The second term on the right-hand side is always positive, so the active–passive particle diffuses faster than a purely Brownian one, yet always slower than a fully active particle, since it never exceeds $s^2/2D_r$. In this sense, the two-state model straddles both limits: more diffusive than the passive case, less diffusive than the active one.

The model also connects to the run-and-tumble particle (RTP) dynamics familiar from motile bacteria such as \emph{E.~coli}~\cite{Schnitzer_1993}. An RTP swims at constant speed $s$ until a sudden reorientation (a tumble) occurs, with rate $\alpha_0$. With translational and rotational diffusion included, its effective diffusivity is~\cite{Cates_2013}
\begin{equation}
D_{eff} = D_t + \frac{s^2}{2(D_r+\alpha_0)}.
\end{equation}
Comparison with Eq.~(\ref{eq: effective diffusion coefficient}) shows that the two expressions can be matched by suitable choices of $\alpha$ and $\beta$. Concretely, one must solve
\begin{equation}
\frac{\alpha}{(\alpha+\beta)^2} \left( \alpha+ \beta \frac{D_r}{\alpha+\beta+D_r}\right)
= \frac{D_r}{D_r+\alpha_0}.
\end{equation}
All parameters are nonnegative. For fixed $D_r$ and $\alpha_0$, this defines an implicit relation $f(\alpha,\beta)=0$. For $\alpha>0$, $f$ is continuously differentiable with $\partial f/\partial\beta \neq 0$, so by the implicit function theorem~\cite{Krantz_2013} there exists a unique function $\beta=g(\alpha)$. In this way, the RTP can be mapped onto an active–passive Brownian particle, at least at the level of long-time diffusivity. A stricter equivalence between the two models, however, remains unproven.

\section{Summary}

We have analyzed the motion of an active particle whose self-propulsion speed $v(t)$ switches stochastically between two states: an \emph{active} state with speed $s$ and a \emph{passive} state with zero speed. We have called this the \emph{active–passive Brownian particle}, to emphasize its dual nature. The motivation for such a model comes from a simple observation: many motile biological entities do not move continuously, but instead turn their activity on and off in a seemingly random manner.  

By calculating the first two moments of the displacement, we have characterized the path statistics of this motion. The results reveal that an active–passive Brownian particle diffuses more rapidly than a purely passive Brownian particle, but more slowly than an active Brownian particle with the same parameters.  

Finally, we have shown that a run-and-tumble particle can be mapped onto the active–passive Brownian particle by an appropriate choice of $\alpha$ and $\beta$, so that both share the same effective diffusion coefficient. In this sense, the active–passive model serves as a compact, analytically tractable framework that interpolates between purely passive, purely active, and run-and-tumble dynamics.

\section*{Acknowledgment}

I am deeply indebted to my parents, whose support and encouragement over the years have made possible whatever I have managed to accomplish. For this, and for much more, I thank them.

\section*{Appendices}
\appendix


\section{Random Telegraph Process}
\label{sec: Random Telegraph Process}

The two-state process in Eq.~(\ref{eq: v(t)}) tells us how the propulsion speed $v(t)$ evolves in time. It is the classic \emph{random telegraph process}—a discrete-state Markov process with only two allowed values: $v=0$ (passive) and $v=s$ (active).  

The simplest way to characterize it is through its master equations. Let $p(0,t|v_0,t_0)$ be the probability of finding the particle passive at time $t$ given $v(t_0)=v_0$, and similarly for $p(s,t|v_0,t_0)$. These satisfy
\begin{align}
\label{eq: first master equation}
\frac{\partial}{\partial t} p(0,t|v_0,t_0) &= \beta\,p(s,t|v_0,t_0)-\alpha\,p(0,t|v_0,t_0),\\ 
\label{eq: second master equation}
\frac{\partial}{\partial t} p(s,t|v_0,t_0) &= \alpha\,p(0,t|v_0,t_0)-\beta\,p(s,t|v_0,t_0),
\end{align}
with $\alpha$ the activation rate and $\beta$ the deactivation rate. The two equations are not independent—adding them simply yields $p(0,t|v_0,t_0)+p(s,t|v_0,t_0)=1$.  

Solving Eqs.~(\ref{eq: first master equation})–(\ref{eq: second master equation}) gives
\begin{align}
\label{eq: first conditional prob}
p(0,t|v_0,t_0) &= \frac{\beta}{\alpha+\beta} + \left(\delta_{v_0,0}-\frac{\beta}{\alpha+\beta} \right) e^{-(\alpha+\beta)(t-t_0)},\\ 
\label{eq: second conditional prob}
p(s,t|v_0,t_0) &= \frac{\alpha}{\alpha+\beta} + \left(\delta_{v_0,s}-\frac{\alpha}{\alpha+\beta} \right) e^{-(\alpha+\beta)(t-t_0)},
\end{align}
with $\delta$ the Kronecker delta. For our purposes, we set $t_0=0$.  

From these probabilities, the averages we need follow almost immediately. The mean speed is
\begin{align}
\label{eq: average of v(t)}
\langle v(t) \rangle = s \,p(s,t|v_0,0),
\end{align}
and the two-time correlation, for $t_2>t_1$, is
\begin{align}
\label{eq: average of v(t1)v(t2)}
\langle v(t_1) v(t_2) \rangle = s^2 \,p(s,t_2|s,t_1)\,p(s,t_1|v_0,0),
\end{align}
where $p(s,t_2|s,t_1)$ has the same functional form as Eq.~(\ref{eq: second conditional prob}). For $t_1>t_2$, we simply exchange $t_1$ and $t_2$.


\section{Angular Ensemble Averages }
\label{sec: Angular Ensemble Averages}

Before solving the translational Langevin equation~(\ref{eq: translational Langevin eq}), we need the orientation statistics. With $\eta(t)$ Gaussian white noise, $\varphi(t)$ is a Gaussian Markov process. Its first two moments follow immediately from Eq.~(\ref{eq: rotational Langevin eq}):  
\[
\langle \varphi(t) -\varphi_0 \rangle  = 0, \quad \langle (\varphi(t) -\varphi_0)^2 \rangle = 2D_r(t-t_0),
\]
where $\varphi_0$ is the initial angle.  

The corresponding one-point conditional probability density is therefore
\begin{equation}
\label{eq: one-joint probability density}
 p(\varphi,t) = \frac{1}{\sqrt{4\pi D_r (t-t_0)}} \exp\left(-\frac{(\varphi-\varphi_0)^2}{4D_r (t-t_0)}\right),
\end{equation}
which we can also write as $p(\varphi,t|\varphi_0,t_0)$.  

Because $\varphi(t)$ is Markovian, any joint distribution can be built from the transition probabilities $p(\varphi_2,t_2|\varphi_1,t_1)$. For a continuous-state Markov process $X(t)$, conditional averages are defined by
\begin{equation}
\label{eq: first moment}
\langle f(X(t)) | X(t^\prime)=x^\prime \rangle = \int_{-\infty}^{\infty} dx \,f(x) \,p(x,t|x^\prime,t^\prime), \quad t\ge t^\prime,
\end{equation} 
and
\begin{equation}
\label{eq: second moment}
\langle g(X(t_1),X(t_2)) | X(t^\prime)=x^\prime \rangle = \int_{-\infty}^{\infty} dx_1 \int_{-\infty}^{\infty} dx_2\,g(x_1,x_2) \,p(x_2,t_2|x_1,t_1) \,p(x_1,t_1|x^\prime,t^\prime),
\end{equation} 
valid for $t_2 \ge t_1 \ge t^\prime$. (For discrete-state processes, the integrals become sums.)  

The quantities we require are $\langle \cos\varphi(t)\rangle$, $\langle \sin\varphi(t)\rangle$, $\langle \cos\varphi(t_1) \cos\varphi(t_2) \rangle$, and $\langle \sin\varphi(t_1) \sin\varphi(t_2) \rangle$. Using Eqs.~(\ref{eq: first moment})–(\ref{eq: second moment}) with the Gaussian transition probability (\ref{eq: one-joint probability density}), we obtain
\begin{align}
\label{eq: average of cos1}
\nonumber
\langle \cos\varphi(t) \rangle &= \int_{-\infty}^{\infty} d\varphi \,\cos\varphi \, p(\varphi,t|\varphi_0,0) =\cos\varphi_0\,e^{-D_r t}, \\
\langle \sin\varphi(t) \rangle &= \int_{-\infty}^{\infty} d\varphi \,\sin\varphi \, p(\varphi,t|\varphi_0,0) =\sin\varphi_0\,e^{-D_r t}.
\end{align} 
and
\begin{align}
\label{eq: average of cos1cos2}
\nonumber
\langle \cos\varphi(t_1) \cos\varphi(t_2) \rangle &= \int_{-\infty}^{\infty} d\varphi_1 \int_{-\infty}^{\infty} d\varphi_2\,\cos\varphi_1 \cos\varphi_2 \,p(\varphi_2,t_2|\varphi_1,t_1) \,p(\varphi_1,t_1|\varphi_0,0) \\
\nonumber
&=\frac{1}{2} \left[ 1+\cos 2\varphi_0\,e^{-4D_r t_1}\right] e^{-D_r(t_2-t_1)}, \quad t_2\ge t_1, \\ 
\langle \sin\varphi(t_1) \sin\varphi(t_2) \rangle &= \int_{-\infty}^{\infty} d\varphi_1 \int_{-\infty}^{\infty} d\varphi_2\,\sin\varphi_1 \sin\varphi_2 \,p(\varphi_2,t_2|\varphi_1,t_1) \,p(\varphi_1,t_1|\varphi_0,0) \\
&=\frac{1}{2} \left[ 1-\cos 2\varphi_0\,e^{-4D_r t_1}\right] e^{-D_r(t_2-t_1)}, \quad t_2\ge t_1.    
\end{align}

\newpage

\bibliography{Refs}
\bibliographystyle{unsrt}

\end{document}